\documentstyle[prb,preprint,aps]{revtex}

\begin{document}

\begin{center}

{\large\bf Low energy excitations in crystalline perovskite oxides:
Evidence from noise experiments}
          
\vspace{0.75cm}
Arindam Ghosh\footnote[1]{email: aghosh@physics.iisc.
ernet.in} and A. K. Raychaudhuri\footnote[2]{On lien to National 
Physical Laboratory, New Delhi 110012, India.}

\vspace{0.25cm}
Department of Physics\\ Indian Institute of Science, Bangalore 560 012,
India\\

\vspace{0.75cm}
R. Sreekala, M.Rajeswari and T. Venkatesan
      
\vspace{0.25cm}
Center for superconductivity research\\ University of Maryland, College
Park, MD 20742 ,USA\\
                  
\end{center}

\vspace{1.0cm}
\begin{abstract}
In this paper we report measurements of 1/f noise in a crystalline metallic 
oxide with perovskite structure down to 4.2K. The results show existence of
localized excitations with average activation energy $\approx$ 70-80 meV which 
produce peak in the noise at T $\approx$ 35-40K. In addition, it shows clear 
evidence of tunnelling type two-level-systems (as in glasses) which show up in
noise measurements below 30K. 
\end{abstract}
\newpage

A number of solids exhibit excitations which are of much lower energy
than the characteristic energy scale of typical lattice excitations.
Most of the time the low energy excitation arises from defect motions 
where atoms or group of atoms have more than one equilibrium configurations 
of almost equivalent energy separated by shallow barriers which are accessible 
by tunneling or thermal activation. A particular type of such defect motions 
are atomic or molecular tunneling of defects seen in a number of doped 
crystalline systems~\cite{VNP,SKW}. In case of glasses and other amorphous
solids the tunneling centers are of a special kind(popularly called Two Level 
Systems) which  give rise to the Universal low temperature properties of 
glasses~\cite{HAR}. This particular field has been well studied by a number 
of techniques. Recently, very sensitive torsional oscillator experiments have 
been used to study the evolution of low energy excitations in amorphous 
silicon~\cite{BWP}. It has also been known for some time that certain 
crystalline solids including metals and metallic alloys show glass-like low 
energy excitations~\cite{VRP}. In this communication we report existence of 
low energy excitations in three dimensional cubic  perovskite oxides 
which were observed through sensitive noise experiments.

It has been known since a decade that the cuprate superconductors (belonging 
to perovskite class) and perovskites with ABO$_3$ structure  can posses 
"glass-like" low energy excitation originating from oxygen defects~\cite{ZEF}. 
In recent years, as the quality of the high T$_c$ crystals and films improved, 
the observed magnitude of these excitations have become less, often falling far 
short of the detectability of the techniques used, like specific heat.   

We have undertaken this experiment to investigate the following issues: (i) to 
establish that existence of low energy excitations may be generic in ABO$_3$ 
type perovskite oxides and (ii) to establish that noise experiments can detect 
such excitations which cannot be detected by other techniques. In particular, 
the existence of low energy excitations in epitaxial films of perovskite oxide 
will have implications on the transport properties and
is a new addition to our knowledge of oxygen dynamics of these materials. 
To our knowledge this is the first experimental determination of noise in
normal conducting perovskite oxides down to low enough temperature so
that the low energy excitations can be detected.

The choice of the material for our study is LaNiO$_{3-\delta}$. This
is a perovskite oxide with almost cubic structure~\cite{RSR}. 
It is electrically
conducting in stoichiometric form ($\delta \approx$ 0). A number of
experiments (including noise experiments) show that the oxygen is
highly mobile in this oxides when $\delta \approx$ 0. Under normal
pressure the oxygen diffuses out of the system even at room
temperature and $\delta$ increases as observed recently 
through sensitive noise experiments~\cite{AGR}.

Electronic transport in this system is strongly linked to oxygen
stoichiometry($\delta$) and resistivity($\rho$) increases sharply as
$\delta$ increases~\cite{NGR}. This makes electronic conduction noise 
a very sensitive probe of oxygen dynamics in this solid. In a previous 
study~\cite{AGR} we have shown that noise experiments can detect long
range oxygen diffusion for T>150K. In this paper we show that there
exists clear evidence of low energy excitations in these solids which
most likely arises from localized dynamics of "frozen defects" of the
oxygen lattice at lower temperatures (T<100K). It is not necessary
however, that these excitations can be called "glass like". 

Epitaxial films of LaNiO$_{3-\delta}$ with nominal thickness $\approx$ 150 nm
were deposited on LaAlO$_3$ substrates by pulsed laser deposition
(PLD) using excimer laser. The details of film preparation is given 
elsewhere~\cite{MSG}. The epitaxy of the films are tested by
X-ray diffraction. The resistivity of the film was measured by a four
probe technique over the temperature range 4.2K<T<380K. The film
used by us had $\rho_{RT} \approx$ 0.6 m$\Omega$-cm and $\rho_{4.2K}
\approx$ 0.25 m$\Omega$-cm and an estimated $\delta
\approx$ 0.05. This is close to but somewhat larger than the
resistivity obtained in best bulk samples prepared by high pressure
oxygen treatment~\cite{XPL}. 

The noise was measured by a five probe ac technique (carrier
frequency 377 Hz) using samples of bridge type configuration with
active volume for noise detection ($\Omega$) $\approx$ 5$\times$
10$^{-8}$ cm$^3$ with peak current density $\sim$ 10$^{4}$ A/cm$^2$.
The noise determination used digital signal processing techniques
described elsewhere~\cite{JS}. The temperature control during noise
determination was better than 10mK. The background noise primarily
consisted of Johnson noise 4k$_B$TR from the sample.
Quadratic dependence of the spectral power density 
S$_v$(f) on V$_{bias}$ ensured ohmic character of the noise.  
 
To normalize our data taken at different frequencies we use a
quantity($\gamma$/n) defined as~\cite{HKV}:

\begin{equation}
$$\gamma$/n = f$\Omega$S$_v$(f)/V$^2_{bias}$$
\end{equation}

\noindent where n is the carrier density, S$_v$(f) is the spectral
power density at frequency f measured with bias V$_{bias}$. This is 
justifiable given the fact that S$_v$(f) $\propto$ V$_{bias}^2$ and
it has nearly 1/f dependence. [In general S$_v$(f) has 1/f$^\alpha$
dependence with the exponent close to 1] We express our results as $\gamma$/n 
because there are uncertainties in exact value of the carrier 
concentration n.Such uncertainties however do not affect the main 
conclusions of this work. The 
quantities on the right hand side are all experimentally measurable. 
In fig.1 we show the normalized power ($\gamma$/n) as a function of
temperature at three frequencies. A few plots of typical noise spectra 
are shown in fig.2a. The spectra are shown as S$_v$(f) vs. f to accentuate
the deviation of $\alpha$ from 1. Beyond 60K noise increases monotonically 
with T. The rapid increase of
$\gamma$/n above 150K is associated with long range oxygen diffusion.
This particular issue has been discussed previously~\cite{AGR}. 
In this brief report we focus the region T<60K where the long range 
diffusion has ceased.

Below 60K there are distinct peaks in noise power as a function of T. 
The peak positions as well as the peak heights
of both the peaks depend on the measuring frequencies. The peak
positions shift to higher temperatures with increasing frequency
which we interpret as a signature of thermally activated process. 
From fig.2 we can also see that there is a perceptible change in 
$\alpha$ as we cross the peak region. 

If a thermally activated process with a characteristic time scale
$\tau$ governs the underlying dynamics associated with the distribution 
of the electron scattering centers, it can give rise to
excess noise in the frequency scale $\sim$1/$\tau$. 
Dutta-Horn model envisages such thermally activated
centers as the origin of 1/f$^\alpha$ noise provided the activation energy 
has a broad distribution with respect to k$_B$T~\cite{DH}. 
As we see from fig.1 for T<35K the noise magnitude decreases rather 
slowly with decreasing temperature which requires a broad ($\gg$kT) 
distribution of activation energies at these temperatures. It is this 
observation that has prompted us to use the Dutta-Horn theory to extract the 
density of states(DOS) D(E) in the dynamics of the scatterers from 
the observed temperature variation of $\gamma$/n. In the framework
of the Dutta-Horn model, the energy E is related to $\tau$ 
through the relation $\tau$ = $\tau_0$ exp[E/k$_B$T], 
$\tau_0^{-1}$ being the attempt frequency. The value of $\tau_0$
($\approx$ 3$\times$10$^{-11}$ sec) was obtained from the shift in 
peak temperatures for different observation frequencies.
We show in fig.3 the result of calculation
of D(E) based on this model. In this model E $\approx$
-k$_B$Tln($\omega\tau_0$) where $\omega$ = 2$\pi$f is the measuring 
frequency. We find that the
measurements of $\gamma$/n for all the three frequencies give D(E)
that peak around 70$\pm$4 meV and 84$\pm$10 meV.  
To check the self-consistency of the analysis we calculated the 
exponent ($\alpha$) at 5Hz from the observed spectra at different 
temperatures and compared the values with that predicted by the 
Dutta-Horn model (fig.2b). We used the relation 
$\alpha$ = 1 - ($\partial$lnS/$\partial$lnT - 1)/ln($\omega\tau_0$).
$\partial$lnS/$\partial$lnT was calculated from a smooth curve fit to 
the experimental data. It is cear from the figure that observed 
$\alpha$ agrees well with the prediction. Thus the analysis of the 
spectra in the framework of the Dutta-Horn model points to the existence 
of a broad flat spectrum of low energy relaxing states accompanied 
by two distinct peaks.
   
LaNiO$_3$ belongs to ABO$_3$ class of perovskite oxides in which
the oxygen is the mobile species and defect chemistry plays an 
important role~\cite{CIC}. 
The oxygen diffusion in this material is
mediated by defects like oxygen vacancy and has an activation energy
$\sim$ 1eV~\cite{AGR}. In the temperature range of
our interest the long range diffusion is frozen. Only kinetics that
seems to be allowed is local relaxation involving oxygen and
oxygen-vacancy combination. The important question however is that
the observed average activation energy $\langle$E$\rangle$ for this
process is $\approx$ 70 meV which is more than an order of magnitude 
less than
that seen for long range diffusion. Typical $\Theta_D$ of this oxide
found from specific heat as well as Bloch-Gruneissen theory is
$\approx$ 350K - 400K which shows the scale of typical lattice vibration
to be smaller than the activation energy associated with the peaks. 
Thus the activation energy seen by us has an energy scale in between 
the lattice vibration scale and the activation energy associated with 
long range diffusion. Also there seems to be a split
peak in D(E) with activation energy differing by approximately 10 meV
implying that there are two classes of oxygen 
defect sites with slightly differing activation energy. Two such different 
sites can arise if the structure is slightly distorted from the cubic 
structure.These small distortions from cubic structure do exist in these 
oxides~\cite{RSR}. Our noise experiments thus clearly show 
the existence of low energy localized excitation in this solid. In
the following we will investigate the region with E $\leq$ 60 meV.

We find from the D(E) vs. E curve
that while D(E) $\rightarrow$ 0 for E $\geq$ 100 meV, there are
substantial number of states left with much lower energy and D(E)
$\rightarrow$ Constant for E $\leq$ 60 meV. 
It raises the probability of existence of a
glass-like excitations in these crystalline solids with nearly 
constant density of states. To investigate
this point further we continued the measurement down to T $\approx$
4.2K. We find that below 30K the magnitude of the noise is $\propto$
T as seen in the inset of fig.1. At T < 30K, defect relaxation 
by thermally activated processes         
freeze out because the relaxation time $\tau$ becomes very large. 
A likely mechanism that can give rise to such low energy
processes at low temperature is tunnelling of atomic or molecular
species as were found in certain doped alkali halide systems~\cite{VNP}. 
It has
been shown before that in presence of random strain fields such two level
tunnelling defects can broaden out to give tunnelling states with a 
broad energy spectrum~\cite{SKW,NOTE1}.
It has been shown theoretically that the temperature dependent relaxation 
rates of the tunneling states (arising from interactions with both 
electrons and phonons)
lead to fluctuations in the number density of these states~\cite{KRE}. 

It is known that the tunneling systems in glasses can interact with the 
electrons and thus provide an extra scattering mechanism. The fluctuation
in the equillibrium number density of the tunneling states thus gives rise
to a fluctuation in the electron scattering rate leading to noise in the 
electrical resistivity. The noise power spectrum of this process measured 
over a volume $\Omega$ is given as~\cite{KRE},

\begin{equation}
\frac{S_{v}(f)}{V^2} = \frac{\langle C^2 \rangle}{n_{imp}^2}
                       \frac{\pi D_{TLS}}{\Omega}
                       \frac{k_{B}T}{\omega}
\end{equation}

\noindent where D$_{TLS}$ is the constant density of states of the
TLS, $\omega$ = 2$\pi$f, C is a dimesionless coupling constant 
$\sim$ 10$^{-2}$ and n$_{imp}$ is the concentration of impurities 
which scatter electrons. This will give rise to a normalized spectral
power 

\begin{equation}
\frac{\gamma}{n} = \frac{1}{2}\langle C^2 \rangle k_BT
                   \frac{D_{TLS}}{n_{imp}^2}
\end{equation}

Our experimental observation for T < 30K (inset of fig 3) 
show a clear linear
dependence on T as mentioned before and the power spectra are also
1/f type. This we consider as an evidence for existence of glass-like 
low energy excitations in these oxides. Taking n$_{imp}$ as the 
number density of oxygen vaccancies (our sample has $\delta \approx$ 
0.05) we obtain D$_{TLS} \sim$ 10$^{35}$ erg$^{-1}$cm$^{-3}$. This
assumes every impurity atom is associated with a TLS. However, in
glasses and low doped alkali halides it is known that the TLS arises 
from a group of impurity atoms and n$_{imp} \ll$ oxygen defect density. 
In that case D$_{TLS} \ll$ 10$^{35}$ erg$^{-1}$cm$^{-3}$. A D$_{TLS} 
\approx$ 10$^{31}$-10$^{33}$ erg$^{-1}$cm$^{-3}$ is found in oxide
glasses~\cite{HAR}. 

In order to cross-check whether the flatness of the DOS at low energies 
is a necessary 
condition to explain the observed temperature dependence of noise,
we carried out calculations of the noise magnitude as a function of 
temperature with several trial DOS. Assuming an activated scattering 
mechanism we can write~\cite{DH},

\begin{equation}
S(f,T) = \int^\infty_0 D(E) \frac{2\tau_0e^{E/kT}}
                                 {1 + (2\pi f\tau_0)^2e^{2E/kT}}\,dE
\end{equation}

We find that usual bell shaped DOS with width $\gg$ kT does not 
produce the observed temperature dependence of noise particularly at
lower temperatures. Best fit to the data 
was obtained with DOS that have a flat distribution D(E) $\approx$
constant for E < 70 meV. In fig.4 we have shown the calculation based
on two typical choices of DOS as examples. 
The dotted line is calculated from a Lorenzian D(E) having width
of $\approx$ 15 meV and the
solid line is obtained from a Lorenzian with flat and 
finite low energy tail (inset of fig.4). The need for a flat low energy
distribution is clearly visible. Thus our analysis shows that if we 
satisfy two basic requirements we can reproduce our data. These 
two requirements are (a) a collection of TLS with a broad distribution
and (b) a flat DOS at low energy. Apart from these two general aspects 
there is nothing unique about the choice of the DOS.  

In these
perovskite oxides the electronic path is formed by the network of
transition metal and oxygen where the transition metal is at the
center of the octahedron formed by six oxygen ions. The electrical
conduction is strongly modified by any defect in the oxygen lattice.
The noise is thus a very sensitive probe of oxygen
and oxygen-defect dynamics which directly couples to the scattering 
centers which give rise to electrical resistivity and resulting 
1/f$^\alpha$ noise.
The low energy defects seen through the noise experiments arise from 
the oxygen dynamics in the three dimensional transition 
metal-oxygen-transition metal network. All ABO$_3$ oxides 
in particular has such networks. It may thus happen that 
the low energy excitations seen in 
LaNiO$_3$ is generic to all such ABO$_3$ oxides. In high T$_c$ 
cuprates, however, there are different types of chain and plain 
oxygens. It will not therefore be proper to comment on existence of
low energy excitations in cuprates based on measurements on
LaNiO$_3$.

\newpage

{\bf\Large Figure Caption}

\vspace{0.5cm}
{\bf Fig.1.} Temperature dependence of the normalized noise power ($\gamma$/n) 
at low temperature at three measuring frequencies. The solid lines are guide to 
the eye. The inset magnifies the low temperature (T < 35K) tail of the noise
magnitude.

\vspace{0.5cm}
{\bf Fig.2.} (a) Plots of the noise power at four different temperatures. Spectra
at different temperatures are shifted for clarity. (b) Temperature variation of
the spectral slope at 5Hz. The solid line is the prediction of the Dutta-Horn
model obtained from the smooth curve fit to the noise magnitude (see fig.4).

\vspace{0.5cm}
{\bf Fig.3.}  Distribution of activation energy E obtained from the data using 
Dutta-Horn model. Value of $\langle \tau_0 \rangle$ (=3$\times$10$^{11}$ sec)
is obtained from the shifting of the noise peak for different observation 
frequency. The solid lines are guide to the eye.

\vspace{0.5cm}
{\bf Fig.4.}  Calculated noise at 5Hz from trial DOS. The solid line fit to the
data corresponds to the DOS shown in the inset (solid line). The dotted line is 
evaluated from a Lorenzian DOS (for details see text).


\begin{thebibliography}{1-30}

\bibitem{VNP} V. Narayanamurthy and R.O. Pohl, Rev. Mod. Phys. {\bf
42}, 201 (1970)
\bibitem{SKW} Susan K. Watson, Ph.D. thesis, Cornell University
(1992) 
\bibitem{HAR} S. Hunklinger and A.K. Raychaudhuri, Prog. in Low
Temperature Phys., {\bf 9}, 265 (1986)
\bibitem{BWP} X. Liu, B.E. White Jr. R.O. Pohl, E. Iwanizcko, K.M.
Jones, A.H. Mahan, B.N. Nelson, R.S. Crandall, and S. Veprek, Phys.
Rev. Lett. {\bf 78}, 4418 (1997)
\bibitem{VRP} J.E. Vancleve, A.K. Raychaudhuri and R.O. Pohl, Z.
Physik {\bf B93}, 479 (1994) and references cited therein
\bibitem{ZEF} H.E. Zimmer, J. Engert, P. Franch, Z. Henning and E.
Hegenbarth, Ferroelect. letts. {\bf 6}, 33 (1986) and S. Sahling and
J. Sievert, Solid State Communications {\bf 75}, 237 (1990)
\bibitem{RSR} K.P. Rajeev, G.V. Shivashankar and A.K. Raychaudhuri,
Solid State Communications {\bf 79}, 591 (1991)
\bibitem{AGR} Arindam Ghosh, A.K. Raychaudhuri, R. Sreekala, M.
Rajeswari and T. Venkatesan, J. Phys. D: Appl. Phys. {\bf 30}, L75
(1997) 
\bibitem{NGR} N. Gayathri, A.K. Raychaudhuri, X.Q. Xu, J.L. Peng and
R.L. Greene, J. Phys.: Condensed Matter {\bf 10}, 1323 (1998)
\bibitem{JS} J.H. Scofield, Rev. Sci. Instrum. {\bf 58}, 985 (1987)
\bibitem{NOTE1} The possibility of noise originating from a single
fluctuating two level system (TLS) is ruled out since the noise
spectrum at a particular temperature in that case is a Lorenzian~\cite{KRE}
with the frequency exponent $\alpha$ varying from 0 to 2. We, however, 
observed the variation in $\alpha$ typically within 1.00$\pm$0.15 at all
temperatures.
\bibitem{HKV} N. Hooge, T.G.M. Kleinpenning and L.K.J. Vandamme, Rep.
Prog. in Phys. {\bf 44}, 479 (1981)
\bibitem{DH} P. Dutta and P.M. Horn, Rev. Mod. Phys. {\bf 53}, 497
(1981) 
\bibitem{CIC} M. Cherry, M.S. Islam and C.R.A. Catlow, J. Solid State
Chem. {\bf 118}, 125 (1995)\\
M.S. Islam, M. Cherry and C.R.A. Catlow, J. Solid State Chem. {\bf
124}, 230 (1996)
\bibitem{KRE} R. Kree, p285 in "Noise in Physical Systems" (Editor: C.M. 
Van Vlet, World Scientific, Singapore, 1987)
\bibitem{MSG} M. Sagoi, Appl. Phys. Letts. {\bf 62}, 1833 (1993)
\bibitem{XPL} X. Xu, J.L. Peng, Z. Li, H.L. Ju and R.L. Green, Phys. Rev. B 
{\bf 48}, 1112 (1993)


\end{thebibliography}
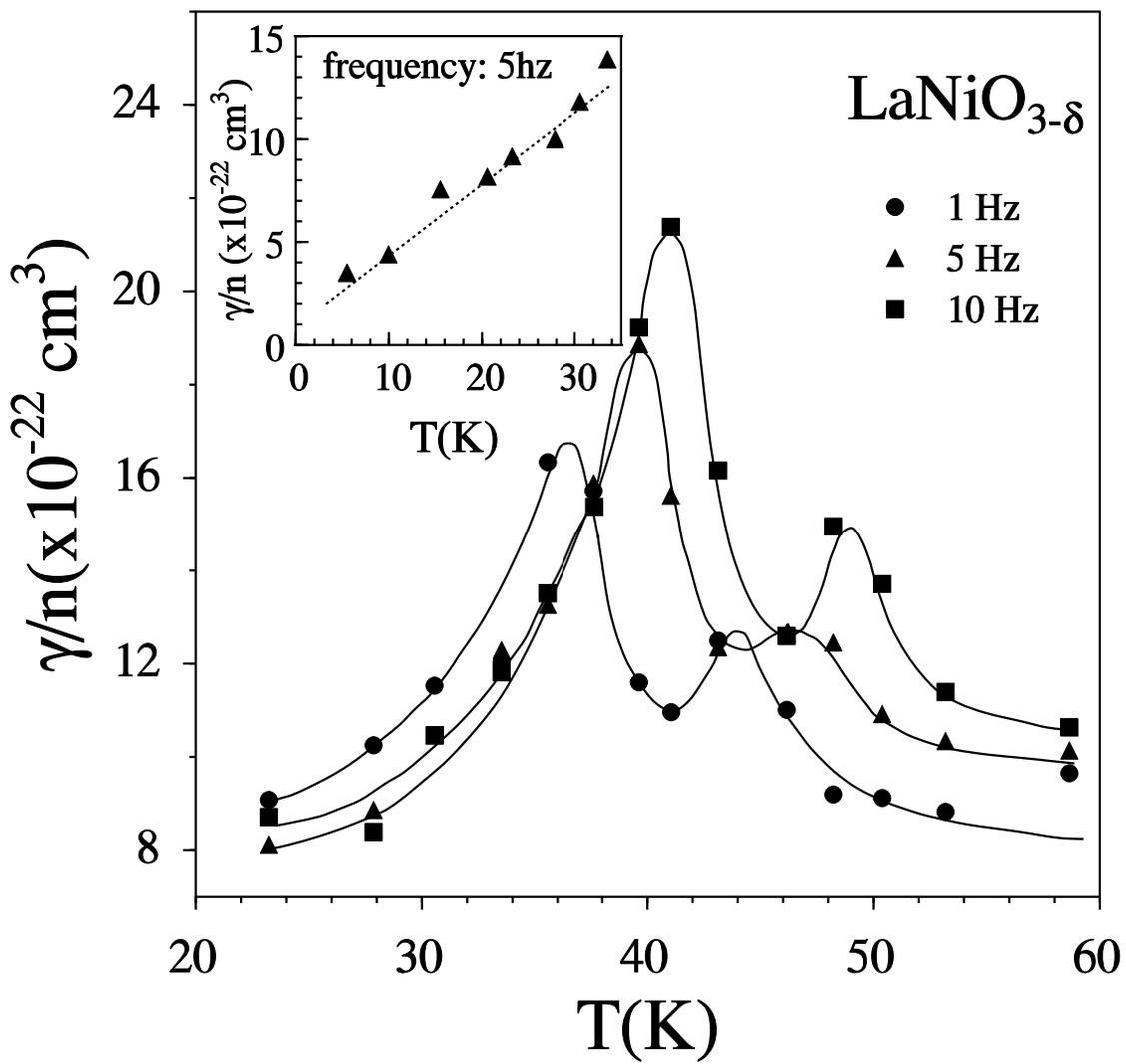
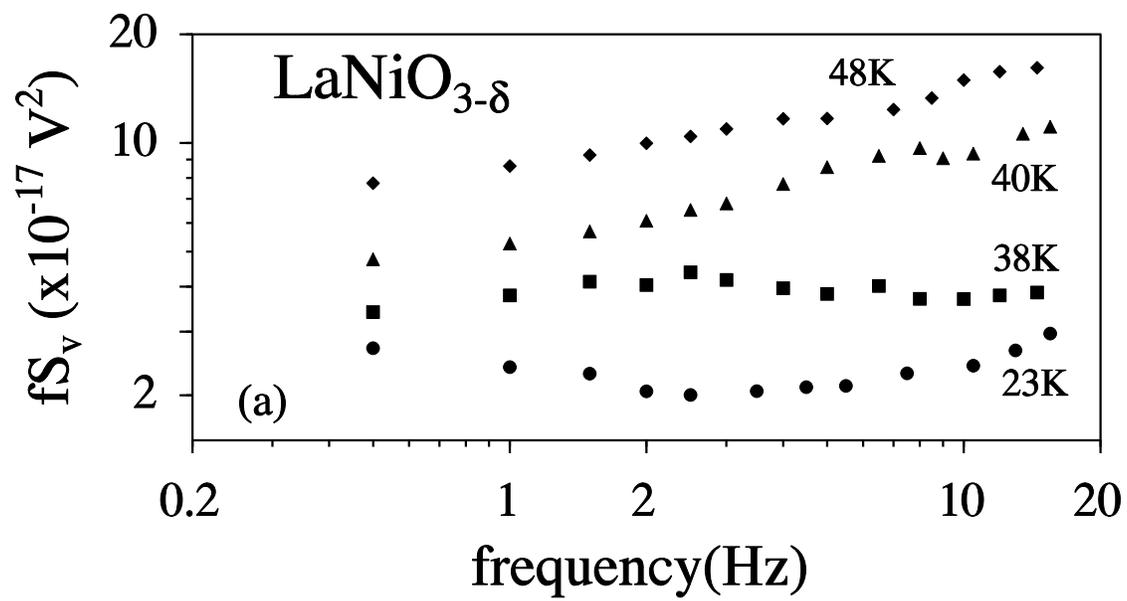
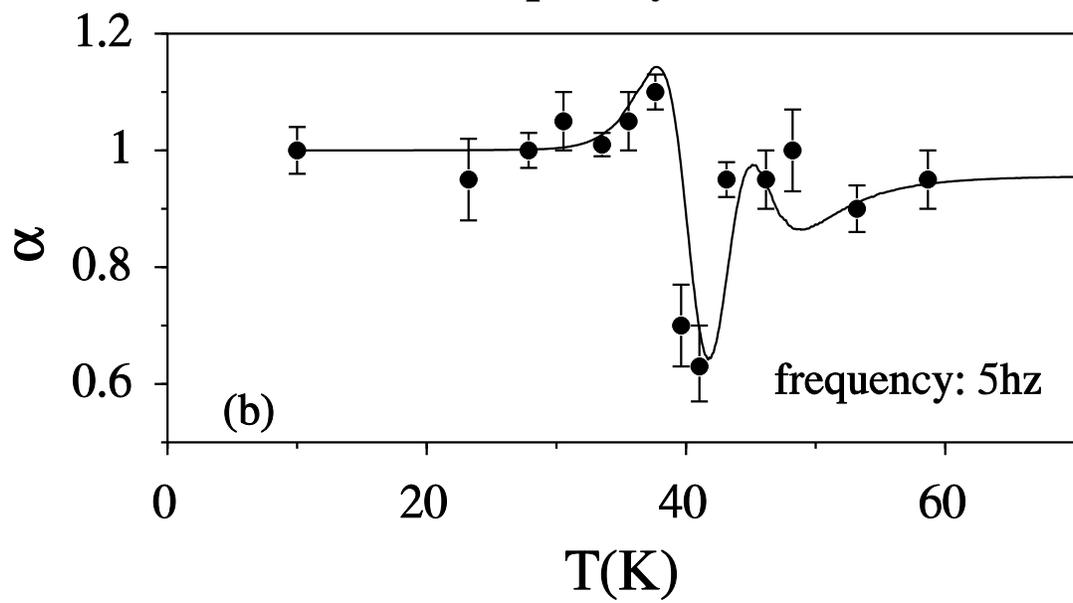
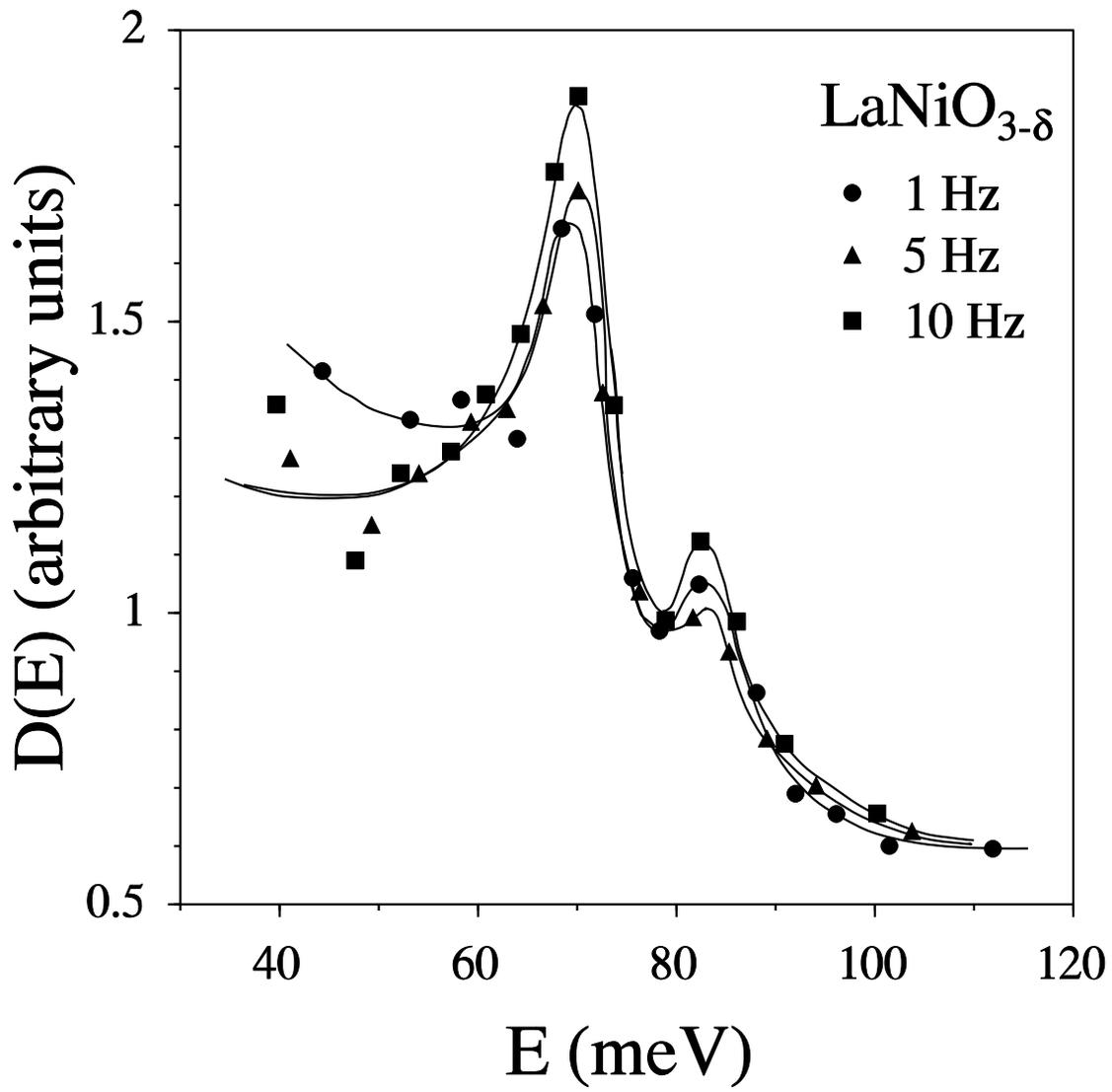
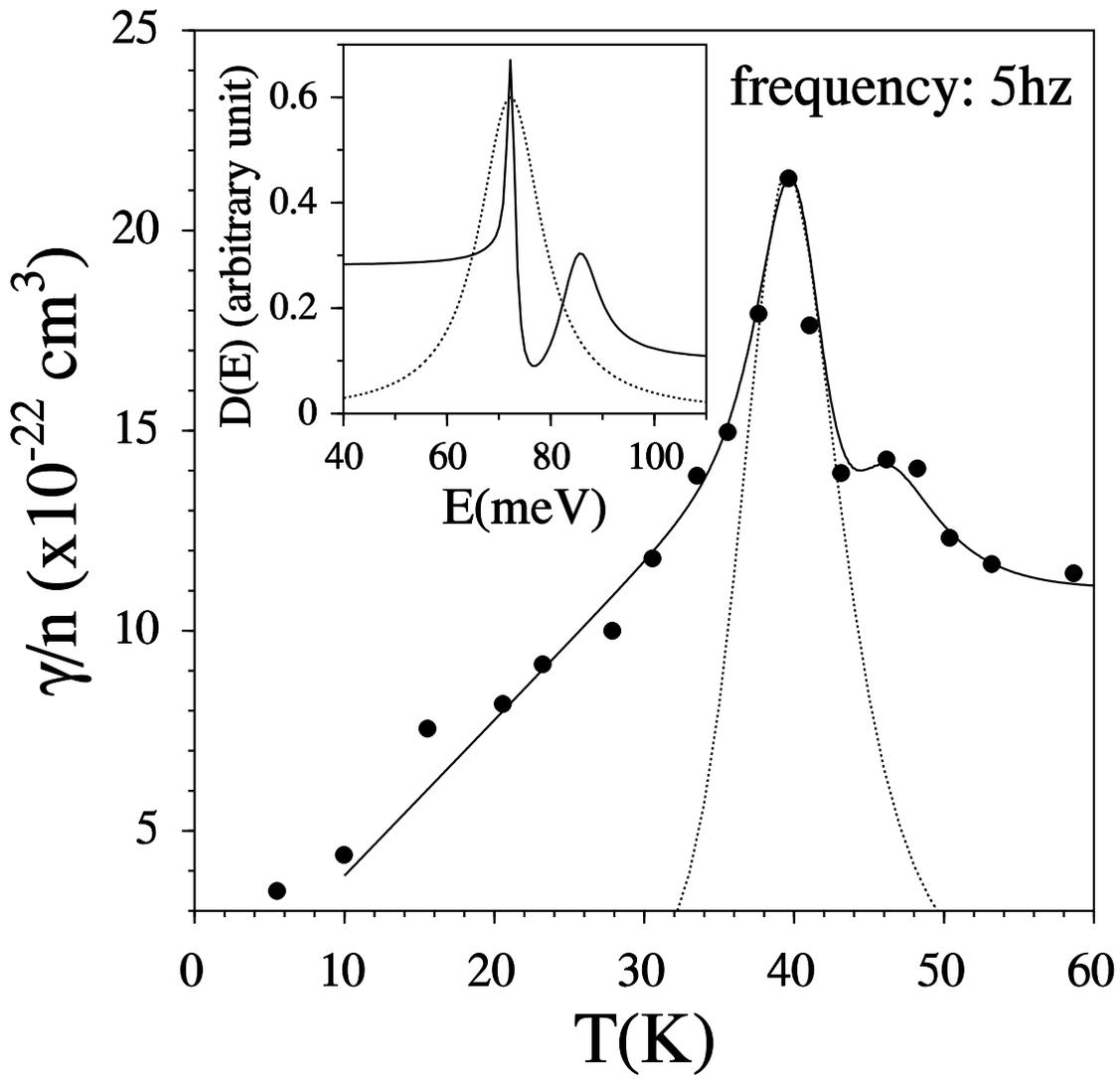
\end{document}